\documentclass[reprint, amssymb, amsmath,aip,cha,]{revtex4-1}

\usepackage{graphicx}
\usepackage{amssymb}
\usepackage{epstopdf}
\usepackage{upgreek}
\usepackage{dcolumn}
\usepackage{bm}
\usepackage[final]{pdfpages}

\expandafter\ifx\csname package@font\endcsname\relax\else
 \expandafter\expandafter
 \expandafter\usepackage
 \expandafter\expandafter
 \expandafter{\csname package@font\endcsname}%
\fi

\newcommand{\be}{\begin{eqnarray}}
\newcommand{\ee}{\end{eqnarray}}
\newcommand{\bfig}{\begin{figure}}
\newcommand{\efig}{\end{figure}}

\newcommand{\crnbs} {Cr$_{1/3}$NbS$_2$}

\begin{document}

\title{Spin Structure of the Anisotropic Helimagnet Cr$_{1/3}$NbS$_2$ in a Magnetic Field}%

\author{Benjamin J. Chapman}
\affiliation{Department of Physics, University of Colorado, Boulder, CO 80309}
\author{Alexander C. Bornstein}
\affiliation{Department of Physics, University of Colorado, Boulder, CO 80309}%
\author{ Nirmal J. Ghimire}
 \altaffiliation[Now at ]{Los Alamos National Laboratory, Los Alamos, NM }
\affiliation{Department of Physics and Astronomy, The University of Tennessee, Knoxville, TN 37996}%
\affiliation{Materials Science and Technology Division, Oak Ridge National Laboratory, Oak Ridge, TN 37831}%
\author{ David Mandrus}
\affiliation{Department of Physics and Astronomy, The University of Tennessee, Knoxville, TN 37996}%
\affiliation{Materials Science and Technology Division, Oak Ridge National Laboratory, Oak Ridge, TN 37831}
\affiliation{Department of Materials Science and Engineering, The University of Tennessee, Knoxville, TN 37996}%
\author{Minhyea Lee}
\email{minhyea.lee@colorado.edu}
\affiliation{Department of Physics, University of Colorado, Boulder, CO 80309}%
\date{\today}

\begin{abstract}
In this letter we describe the ground-state magnetic structure of the highly anisotropic helimagnet \crnbs~ in a magnetic field.  A  Heisenberg spin model with Dyzaloshinkii-Moriya interactions and magnetocrystalline anisotropy allows the ground state spin structure to be calculated for magnetic fields of arbitrary strength and direction. Comparison with magnetization measurements shows excellent agreement with the predicted spin structure.
\end{abstract}

\maketitle
Controlling the electrical properties of materials by manipulating their magnetic structure has been one of the primary themes in the field of  magnetism research and its applications. Major technological innovations have been based on these efforts, such as giant magnetoresistance in magnetic multilayers systems~\cite{Baibich1988} and magnetic tunneling effects~\cite{Moodera1995}.  Recently,  non-trivial spin textures, e.g. solitons~\cite{Togawa2012} and skyrmions~\cite{Muhlbauer2009,Yu:2011}, have received much attention in a similar context~\cite{Fert2013,Romming2013,Bostrem2008,Pappas2012,Togawa2012,Iwasaki2013}. These objects are especially interesting because of the stability granted by their topology.  A detailed understanding of such spin structures, in relation with their effect on electrical properties, is expected to shed light on developing spin-texture based applications~\cite{Iwasaki2013}.

With its layered noncentrosymmetric crystal structure, the helimagnet \crnbs~is well-suited for investigations of spin structure, especially toward controlling electrical properties~\cite{Ghimire2013,Bornstein2014}.  In \crnbs, Cr$^{3+}$ ions are intercalated between the hexagonal $2H$-NbS$_2$ layers and magnetically order at $T_C = 133$ K~\cite{Ghimire2013,Bornstein2014}.  
The crystal structure's lack of inversion symmetry, caused by Cr intercalation, results in a helical magnetic ground state oriented along the crystalline $c$-axis, with spins aligned ferromagnetically within the $ab$ planes.  
Unlike other well known helimagnets with B20 crystal structure~\cite{Muhlbauer2009,Yu:2011}, \crnbs~only breaks inversion symmetry along the $c$-axis, making it ideal for  studying spin-textures in magnetic thin films~\cite{Heinze2011}, which also break inversion symmetry only along the single axis normal to the plane of film.

The quasi 2-dimensional (2D) nature of these layered ferromagnetic planes, paired with strong magnetocrystalline anisotropy, allows a clear distinction between the magnetically hard axis (i.e. $c$-axis of the crystal) and the easy plane ($ab$-plane).  The above qualities of \crnbs~ greatly resemble those of planar magnetic devices fabricated on a substrate, making this material especially attractive as a model system. 

Furthermore,  \crnbs~hosts a chiral soliton lattice phase in the presence of a magnetic field applied within the $ab$-plane~\cite{Togawa2012}.  In this context, solitons are essentially 360$^{\circ}$ domain walls in the spin structure, as illustrated in Fig.~\ref{fig:Mcompare}~(e).  As the field strength increases, the space between adjacent solitons grows~\cite{Dzyaloshinskii1964,Togawa2012}, until a phase transition to a spin-polarized state occurs at $B_c = 0.175$ T~\cite{Bcnote,Bornstein2014}.  Alternatively, if a magnetic field is applied along the $c$-axis (also the helical axis), the transition to a polarized state  occurs at a much larger field of 2.45 T~\cite{Miyadai1983,Ghimire2013},  through a conical state.  A complete description of the spin structure in \crnbs~must unite these disparate magnetic field scales.  It must also describe how the local magnetic structure changes with magnetic fields of varying strength and direction.

In this letter we show that the spin structure of \crnbs~in a magnetic field $\boldsymbol{B}$ is well-described by a 1-dimensional (1D) Heisenberg spin model with Dyzaloshinkskii-Moriya (DM) interactions~\cite{Dzyaloshinsky:1958,Moriya:1960}, a Zeeman interaction, and strong magnetocrystalline anisotropy.  After solving for the ground state of the model and its magnetization, we compare with magnetization measurements of \crnbs.  The predictions follow the data closely, and capture the soliton lattice transition at low magnetic fields applied in the $ab$-plane, the conical transition at high fields applied along the $c$-axis, as well as the behavior at intermediate fields applied at oblique angles.

The inset of Fig.~\ref{fig:Mcompare}~(a) shows a schematic of these measurements, where a magnetic field is applied at an angle $\theta_B$ to the $c$-axis of a \crnbs~sample and the component of the magnetization parallel to that field $M_{\parallel}$ is detected.  In all the measurements, the temperature was held fixed at $T = 4$ K $\ll T_C$, justifying the zero-temperature approximation used in calculations.  
Typical data from these measurements (blue circles) are shown in Fig.~\ref{fig:Mcompare}~(a-d) for different fixed values of $\theta_B$, with the model's predictions plotted as the solid red lines.  The agreement is remarkable.

\begin{figure}[htb]
\begin{center}
\includegraphics[width=0.9\linewidth]{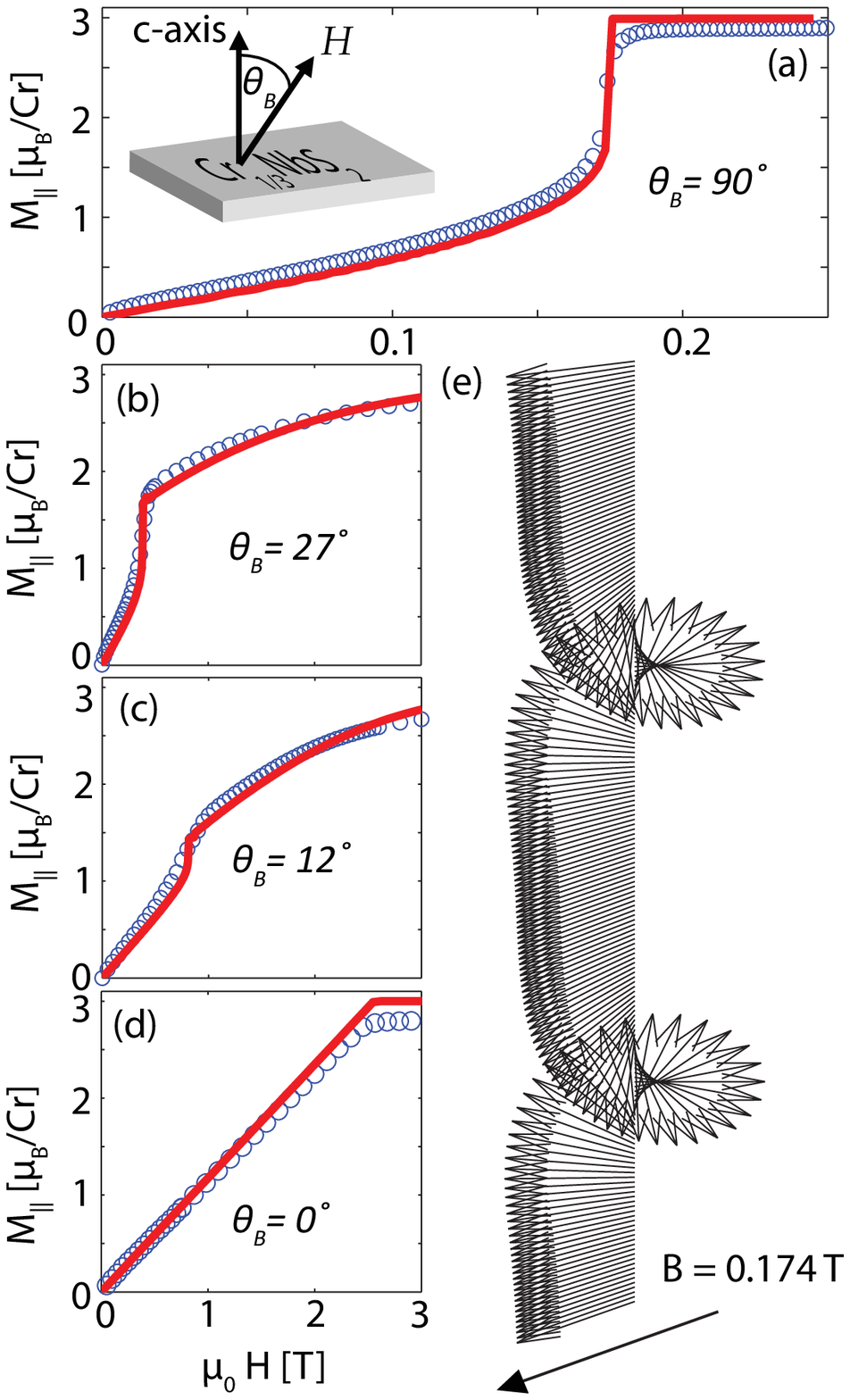}
\caption {\small 
(a-d) Comparison of calculated $M_{\parallel}$ (red lines) with experiments (blue circles) as a function of applied field $H$. A schematic of the measurement is shown in the inset of (a).  $\theta_B$ is fixed    at (a) 90$^{\circ}$, (b) 27$^{\circ}$, (c) 12$^{\circ}$, and  (d) 0$^{\circ}$.   (e) The left-handed spin structure of \crnbs~calculated from Eq.~(\ref{phi}), when $\theta_B = 90^{\circ}$ and $B \to B_c$ from below.  At this field, the magnetic unit cell has grown to more than double its zero-field length $\lambda$.  Two solitons are visible.}
\label{fig:Mcompare}
\end{center}
\efig

We now describe the details of the spin model.  Within a single $ab$-plane, the only spin-spin interaction is exchange, which results in a uniformly polarized state with all spins in the plane aligned.  This allows the low-temperature magnetic structure of \crnbs~ to be studied on a 1D lattice, where sites correspond to local moments of Cr$^{3+}$ ions at different position along the $c$-axis.
The magnetic moment of the spin on the $i^{\mathrm{th}}$ site is represented classically as a 3 dimensional vector of magnitude $s$: $\boldsymbol{s}_i = s \boldsymbol{n}_i$, $\boldsymbol{n}_i$ a unit vector.  The Hamiltonian for the system is
\be
\mathcal{H} =  \sum_{i}{\Big[-J \boldsymbol{s}_i \cdot \boldsymbol{s}_{i+1} - \boldsymbol{D} \cdot (\boldsymbol{s}_i \times \boldsymbol{s}_{i+1})\Big.}\nonumber \\ 
- \Big.\mu_B \boldsymbol{B} \cdot \boldsymbol{s}_i + A \left(\hat{z} \cdot \boldsymbol{s}_i\right)^2\Big],
\label{ham}
\ee
with $J$ and $A$ both positive and $\hat{z}$ along the $c$-axis.  The four terms in the above expression represent the exchange interaction, the DM interaction, a Zeeman interaction, and  magnetocrystalline anisotropy, respectively.  $\mu_B$ is the Bohr magneton.  
In \crnbs,  the crystalline lattice has a 3-fold rotation symmetry about the $c$-axis \cite{Ghimire2013}, so for the DM  term to remain rotationally invariant, $\boldsymbol{D}$ has a non-zero component only along $c$-axis.  We thus put $\boldsymbol{D} = D \hat{z}$. 


The symmetry of this Hamiltonian allows us to confine our study to magnetic field directions  in the $ac$ plane. That is, we fix the azimuthal angle of the magnetic field at $\phi_B = 0$ and examine the ground state at varying field strengths, for different polar angles $\theta_B$ in the interval $[0,\pi/2]$.

To make direct comparisons between the spin-model and measurements of the magnetization, we choose model parameters appropriate for \crnbs. Cr$^{3+}$ ions have  localized spins $S = 3/2$ with magnetic moments of magnitude $\mu_{Cr} \equiv g \mu_B S = 3\mu_B$, consistent with the observed value.  We thus take $s = 3$ in our calculations.  
To fix the remaining parameters $J$, $A$, and $D$ we examine predictions of the model.  At zero field the ground state is helical with pitch $\lambda = 2 \pi a / \delta$, where $a$ is the c-axis lattice constant and $\delta \equiv \arctan(D/J)$. 
The handedness of the helix is determined by the sign of $D$. In \crnbs, $\lambda \approx 40 a$~\cite{Miyadai1983,Ghimire2013} and the helices are left-handed\cite{Togawa2012}, making $\delta \approx D/J = -0.16$ a small parameter.  In magnetic fields of sufficient strength all moments polarize along $\boldsymbol{B}$.  When $\boldsymbol{B}$ is in the $ab$-plane ($\theta_B = 90^{\circ}$), that critical field is $\mu_B B_c = [(\frac{\pi}{4})^2\delta^2 + \mathcal{O}(\delta^3)]Js$ \cite{Izyumov1984,Kishine2005}, while a field along the $c$-axis ($\theta_B = 0$) requires a stronger field of $[(\delta^2 + 2\alpha) + \mathcal{O}(\delta^3)]Js$.  Here $\alpha \equiv A/J$.  $J$ and $A$ may then be determined by comparison with experiments which determine the saturating magnetic fields in these two different field orientations.  This fixes $\alpha \equiv A/J = 0.10$ and $J = 0.2$ meV, the only free parameters in the model.

To efficiently identify the ground state of Eq.~(\ref{ham}) for arbitrary $\boldsymbol{B}$, we make the assumption that all spins have the same polar angle ($\theta$) or $z$-component.  This is a valid assumption for most magnetic fields, and breaks down only when $0<\theta_B\ll \pi/2$, as has been confirmed by direct numerical minimizations of Eq.~(\ref{ham}) where $\theta_i$ varies with $i$.  Further discussion on the validity of this assumption is momentarily deferred.   

In that case, Eq.~(\ref{ham}) may be converted to spherical coordinates. In a continuum limit, the energy per unit length $\mathcal{E}$ is
\be
\mathcal{E} &=& -\frac{Js^2}{La}\int_{-L/2}^{L/2}dz  
\Big[1+\beta_z \cos\theta - \alpha \cos^2\theta  \nonumber \\
&+&\sin^2\theta\Big(-\frac{1}{2} \Big(a\frac{d\phi}{dz}\Big)^2 + \delta a \frac{d\phi}{dz} + \frac{\beta_x}{\sin\theta} \cos\phi\Big) \Big].  
\label{scriptE}
\ee
Here $\phi(z)$ is the azimuthal angle of spin moments, and the dimensionless field $\beta_\nu \equiv \mu_BB_\nu/Js$.  

A stationarity condition on $\mathcal{E}$ for $\phi(z)$ yields a sine-Gordon equation with solution~\cite{Izyumov1984},
\begin{equation}
\phi(z) = \phi_0 - 2\textrm{am}\Big(\sqrt{\frac{\beta_x}{\sin\theta}} \frac{z}{ka} \Big),
\label{phi}
\end{equation} 
where $\phi_0$ is the initial angle, am is the Jacobi amplitude function, and $k$ is an elliptic modulus, chosen to minimize $\mathcal{E}$~\cite{DeGennes1968}.  Substituting this result into Eq.~(\ref{scriptE}) yields
\be
\mathcal{E} &=& -\frac{J s^2}{a} \Big [1 - \alpha \cos^2\theta + \beta_z \cos\theta  \nonumber \\
&+& \beta_x \sin\theta \Big(\frac{2}{k^2} + \frac{\pi \delta}{k K(k)}\sqrt{\frac{\sin\theta}{\beta_x}} - \frac{4 E(k)}{k^2 K(k)} - 1\Big)\Big]
\label{scriptE2}
\ee
where $K$ and $E$ are the complete elliptic integrals of the first and second kind.  In this form, $\mathcal{E}$ can be numerically minimized to determine $\theta$ and $k$.  For that we used an interior-point algorithm~\cite{Byrd1999}.  
This fixes $\phi(z)$ as well as 
the components of the magnetization~\cite{Kishine2014}, 
$M_x 
 = \mu_{\mathrm{Cr}} \sin\theta\left(\frac{2}{k^2} - 1 - \frac{2 E(k)}{k^2 K(k)}\right)$, and 
$M_z = \mu_{\mathrm{Cr}} \cos\theta.$

The red lines shown in Fig.~\ref{fig:Mcompare}~(a-d) are made by combining these results to find $M_{\parallel}$.  Fig.~\ref{fig:Mcolor} shows their variation over the entire parameter space of interest, with $M_x$ in (a), $M_z$ in (b), and $M_{\parallel}$ in (c).
\begin{figure}[htb]
\begin{center}
\includegraphics[width=0.9\linewidth]{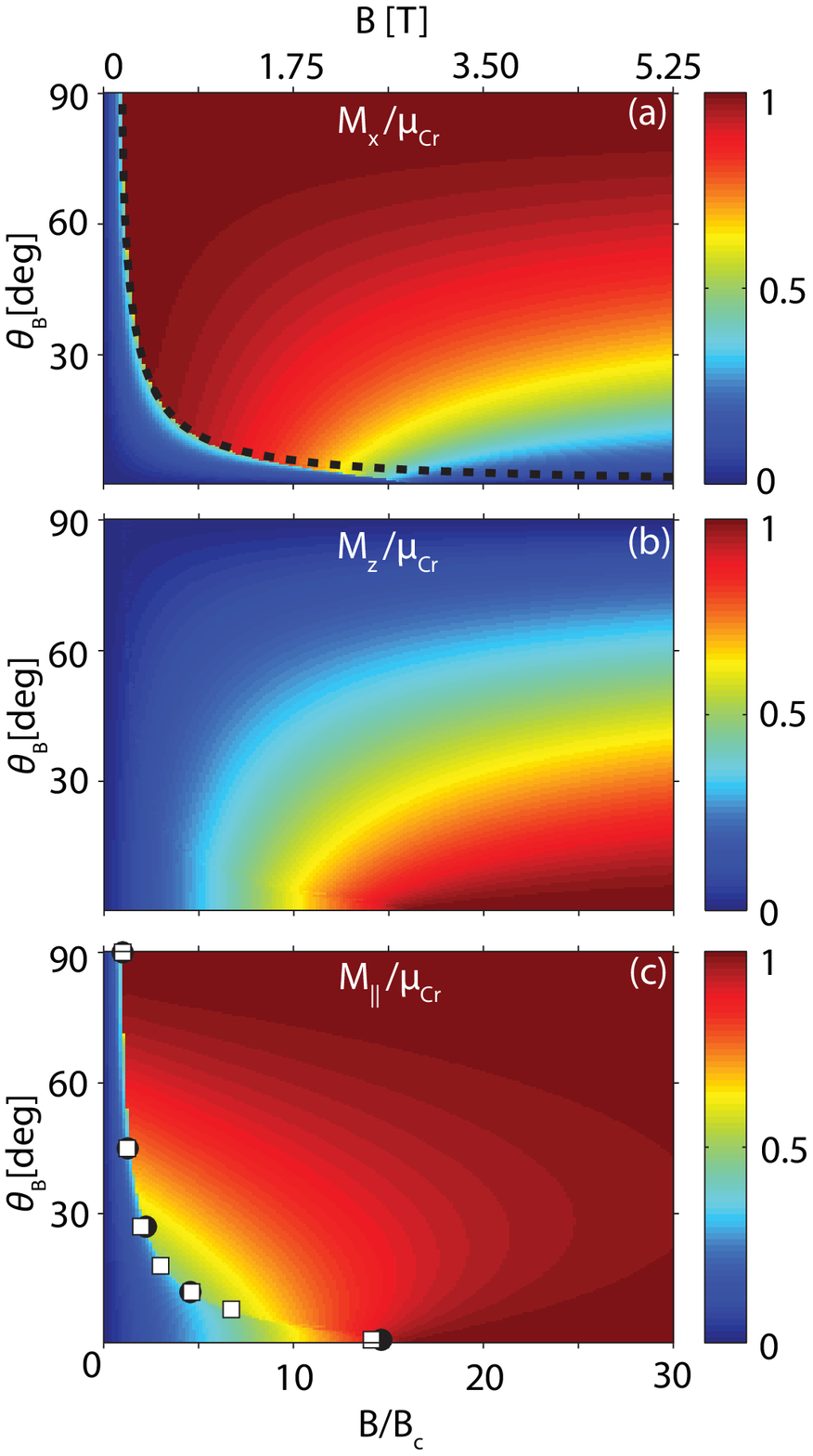}
\caption {\small 
The calculated magnetization normalized by $\mu_{\mathrm{Cr}}$, as a function of field strength and direction: 
(a) $M_x/\mu_{\mathrm{Cr}}$, (b) $M_z/\mu_{\mathrm{Cr}}$ , and (c) $M_{\parallel}/\mu_{\mathrm{Cr}}$.  In (a), the black dashed line indicates an estimate for the field at which the soliton lattice transition occurs: $B_c(\theta_B) \approx B_c / \sin(\theta_B)$.  The markers in (c) represent measurements of $B_c(\theta_B)$ inferred from field sweeps of magnetization (black circles) and magnetoresistance (white squares)~\cite{Bornstein2014}.  Field strengths (x-axes) are given in tesla (top) and normalized by the critical field $B_c(\theta_B = 90^{\circ})$ (bottom).}
\label{fig:Mcolor}
\end{center}
\efig
\appendix

The field for which the length of the magnetic unit cell diverges corresponds to the soliton lattice transition. Its magnitude depends on $\theta_B$, which we write as $B_c(\theta_B)$. In Fig.~\ref{fig:Mcolor}~(a), $B_c(\theta_B)$ is clearly evident in the sudden increase in $M_x$.  Throughout this letter, we denote $B_c(\theta_B = 90^{\circ})$ by $B_c$.  
When $\cos\theta_B \ll 1$, the model predicts a critical field $B_c(\theta_B) \approx B_c / \sin\theta_B$, shown as the dashed black line in Fig.~\ref{fig:Mcolor}~(a).  Only as $\theta_B \to 0$ does the calculated $B_c(\theta_B)$ begin to depart from this estimate.

Note that when $\theta_B < 90^{\circ}$ the magnetization in the $x$ direction grows rapidly through the soliton lattice transition at $B_c(\theta_B)$, but then deceases slowly as $B$, the magnitude of $\boldsymbol{B}$, increases.  This corresponds to a polarization of all moments in (or close to) the $ab$-plane, followed by a slow tilting of the spins out of the plane and along $\boldsymbol{B}$.  This effect is most noticeable when $\boldsymbol{B}$ is near to but distinct from the $c$-axis.  In this regime, $M_x/\mu_{\mathrm{Cr}}$  can approach unity 
at $B =B_c(\theta_B)$, but when $B \gg B_c(\theta_B)$, $M_x(B)/\mu_{\mathrm{Cr}} \to \sin\theta_B \approx \theta_B \ll 1$.

We return now to the discussion of the constant $\theta$ approximation used with Eq.~(\ref{ham}). This is expected to be valid when $B_c(\theta_B) \ll B_c(\theta_B = 0)$, as this means the helix will polarize in the $ab$-plane via a soliton lattice transition before spins develop appreciable components out of the plane.  In \crnbs~this criteria is fairly unrestrictive, as the two fields  of  $B_c(\theta_B)$ and $B_c / \sin\theta_B$ are equal only when $\theta_B \approx \arcsin(\frac{B_c}{B_c(0^{\circ})}) \approx 4^{\circ}$. When the assumption is relaxed and $\theta(z)$ is allowed to vary with $z$, the effect is minimal: as $B$ approaches $B_c$ at a given $\theta_B$, moments whose $x$ components align with $B_x$ tilt slightly downward on average, inclining toward the $ab$-plane, while moments with $x$ components opposite $B_x$ tilt slightly upward. When $\theta_B = 0$, however, the model is again accurate as $\theta(z)$ is constant through the conical transition.


The agreement in $B_c(\theta_B)$ between the model and these measurements is further evidence of electrical transport's sensitivity to the magnetic structure in this material.  The markers in Fig.~\ref{fig:Mcolor}~(c) are measurements of $B_c$ as a function of $\theta_B$ deduced from magnetization (black circles) and $ab$-plane magnetoresistance (white squares) reported previously~\cite{Bornstein2014}.  
Interestingly, $M_{\parallel}/\mu_{Cr}$ is less than one for a wide range of angles $\theta_B$, even when the magnetic field is of the order of several tesla.  This is visible in the broad and shallow dip seen in the center of Fig.~\ref{fig:Mcolor}~(c), and could be relevant to experiments operating in this regime.

In summary we presented a spin model to describe the ground state magnetic structure of \crnbs~in a magnetic field applied in arbitrary direction.  We also detailed an efficient method to calculate that ground state, along with its magnetization.  Comparison with experiments reveals that the model  captures the complex spin structure in this material.  This work supports future efforts to engineer technologies that rely on the manipulation of spin structures in the presence of high magntoanisotropy, like that of \crnbs. 

\vspace{0.1in}
\noindent{\emph {Acknowledgement}} We thank P. Beale and M. Glaser for their insight. 
This work was supported by the US DOE, Basic Energy Sciences, Materials Sciences and Engineering Division (ORNL) and at CU under Award Number DE-SC0006888.
\vspace{0.1in}

\noindent $^{\dagger}$Current Address: Los Alamos National Laboratory, Los Alamos, NM 

%

\end{document}